\documentclass[3p,times,procedia]{elsarticle}
\usepackage{nupha_ecrc}
\usepackage{hyperref}
\volume{00}

\firstpage{1}

\journalname{Nuclear Physics A}

\runauth{}

\jid{nupha}

\jnltitlelogo{Nuclear Physics A}

\usepackage{amssymb}

\usepackage[figuresright]{rotating}

\newcommand{\be}{\begin{eqnarray}}
\newcommand{\ee}{\end{eqnarray}}
\newcommand{\non}{\nonumber\\}

\newcommand{\ave}[1]{\left\langle #1 \right\rangle}

\newcommand{\gev}{{\rm \, GeV}}
\newcommand{\cum}[1]{\kappa_{#1}}

\begin{document}

\begin{frontmatter}



\dochead{XXVIIIth International Conference on Ultrarelativistic Nucleus-Nucleus Collisions\\ (Quark Matter 2019)}

\title{The QCD phase diagram and statistics friendly distributions}


\author[lbnl]{Volker Koch}
\author[krak]{Adam Bzdak}
\author[lbnl]{Dmytro Oliinychenko}
\author[ffm]{Jan Steinheimer}

\address[lbnl]{Nuclear Science Division,
Lawrence Berkeley National Laboratory, 1 Cyclotron Road,
Berkeley, CA 94720, U.S.A.}
\address[krak]{AGH University of Science and Technology,
Faculty of Physics and Applied Computer Science,
30-059 Krak\'ow, Poland}
\address[ffm]{Frankfurt Institute for Advanced Studies, Ruth-Moufang-Str. 1, D-60438 Frankfurt am Main, Germany}

\begin{abstract}
The preliminary  STAR data for proton cumulants for central collisions at $\sqrt{s}=7.7\gev$ are consistent with a
two-component proton multiplicity distribution. We show  that this two-component  distribution is statistics
friendly in that factorial cumulants of surprisingly high orders may be extracted with a relatively
small number of events. As a consequence the two-component model can be tested and verified {\em
  right now} with the presently available STAR data from the first phase of the RHIC beam energy scan. 
\end{abstract}


\end{frontmatter}


\section{Introduction}
\label{sec:intro}
One of the central question of strong interaction research is the possible existence of a
first-order phase transition accompanied by a critical point. This phase transition, if it exists, is
expected to be located in regions of high net-baryon density. Parts of the high density region of the QCD
phase diagram is accessible to experiment through heavy ion collisions at not too high energy. Since
the location of the phase transition and critical point is not known from first principle,
the strategy to search for a phase transition is to scan the high density region of the phase
diagram by measuring relevant observables for a whole range of collision energies. This is one of
the main motivations of the RHIC beam energy scan (BES), the second phase of which (BES-II) 
is presently underway (for a recent review on this topic, see \cite{Bzdak:2019pkr}). One of
the more promising observables for the experimental detection of the QCD phase transition are
fluctuations of conserved charges, most prominently those of the net-baryon number. Near the critical
point (and the phase transition) these fluctuations are predicted to be enhanced \cite{Stephanov:2008qz} which would result in a
non-monotonic behavior of the net baryon cumulants as a function of the collision energy. During the
first phase of the RHIC beam energy scan (BES-I) the STAR collaboration has measured the cumulants of the (net)-proton
distribution for a wide range of energies \cite{Luo:2015ewa,Adam:2020unf}. For the most central
collisions, these data show an
interesting, non-monotonic behavior in the ratio of the fourth order over the second order cumulant
${\cum{4}}/{\cum{2}}$. In addition, this ratio increases to rather large values at the lowest energy
of $\sqrt{s}=7.7\gev$. A further analysis of the STAR data for the lowest energies showed that they
are consistent with very strong, and positive, four-proton as well as sizable, and negative,
three-proton correlations, which both increase in magnitude with decreasing energy
\cite{Bzdak:2016sxg}. While a simple cluster model can describe the magnitude of these correlations
\cite{Bzdak:2016jxo} it fails to get the signs right. In \cite{Bzdak:2018uhv} it was suggested that
the observed correlations are consistent with a two-component or ``bi-modal'' proton multiplicity
distribution consisting of a dominant binomial distribution with mean $\ave{N_{\rm large}}\simeq 40$ and
another small component with  strength $\alpha\simeq 0.3\%$ and mean $\ave{N_{\rm small}}\simeq 25$.
Interestingly, such a distribution is akin to a multiplicity distribution one encounters for a sufficiently small
system in the vicinity of a first order phase transition \cite{Bzdak:2018uhv}. Clearly, additional measurements and
analysis will be needed to verify this hypothesis. Here we will argue that this hypothesis can be
tested {\em right know } with the presently available STAR data.

\section{Two-component model as a statistics friendly distribution}
\label{sec:main}
\begin{figure}[t]
  \begin{center}
  \includegraphics[width=0.37\textwidth]{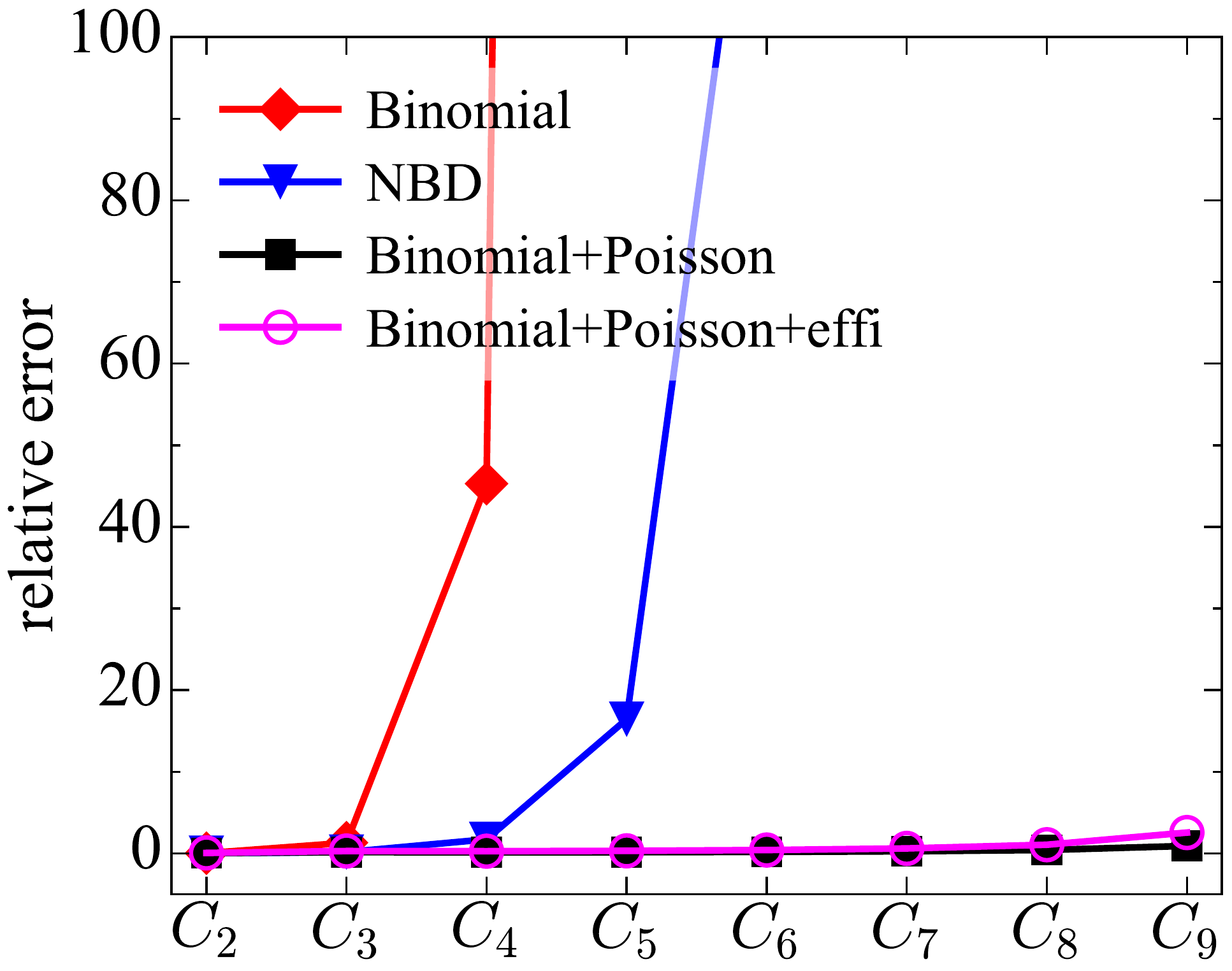}
  \hspace{0.05\textwidth}
  \includegraphics[width=0.46\textwidth]{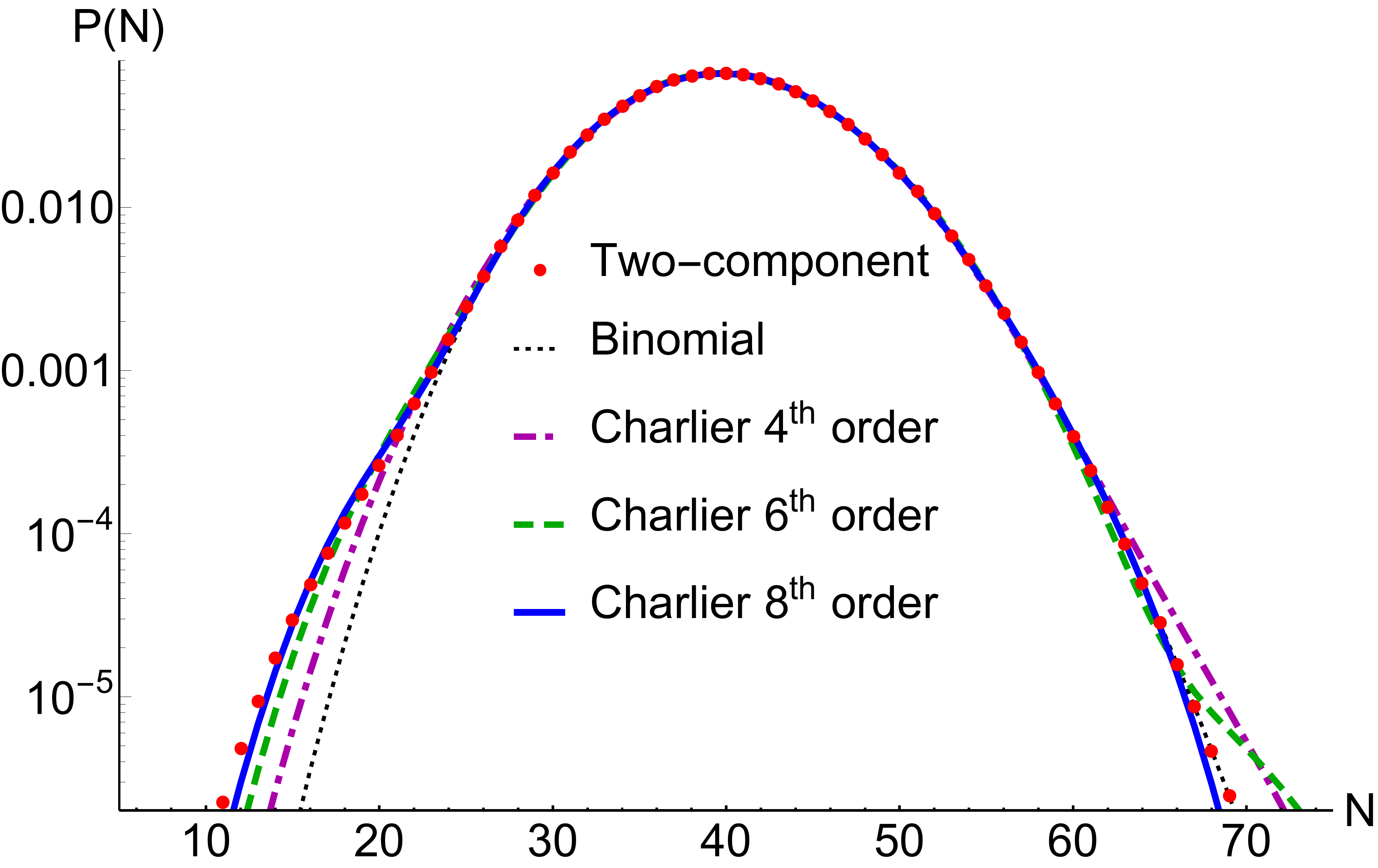}
\end{center}
\par
\vspace{-5mm}
\caption{Left panel: The relative error, $\Delta C_{n}/C_{n}$, of the factorial cumulants for various proton
  multiplicity distributions based on 150k events. The magenta line represents the results if we
  assume that the data are not efficiency corrected. Right panel: Various orders of the
  Poisson-Charlier expansion of the multiplicity distribution compared with that of the
  two-component model, Eq.~(\ref{eq:two_component}), (red points). The black line corresponds to the
  binomial distribution which is obtained from the distribution $P(N)$ of
  Eq.~(\ref{eq:two_component}) by turning off the small component,  $\alpha=0$. 
}
\label{fig:relative_error}
\end{figure}

The two-component distribution, which in Ref.~\cite{Bzdak:2018uhv} was found to reproduce the preliminary
measurement of the proton cumulants  by STAR for the most central collisions at 7.7~GeV \cite{Luo:2015ewa}, is
\begin{equation}
P(N)=(1-\alpha )P_{(a)}(N)+\alpha P_{(b)}(N),
\label{eq:two_component}
\end{equation}
where $\alpha \approx 0.0033$. The distribution  $P_{(a)}(N)$ is given by a binomial ($N_{max}=B=350$,
$p\approx0.1144$) and $P_{(b)}(N)$ is a Poissonian ($\langle N_{(b)} \rangle = 25.3525$). The
distribution $P(N)$ is depicted as the red
points in the right panel of Fig.~\ref{fig:relative_error}.
The four cumulants measured by STAR obviously do not sufficiently constrain this model, which in
itself has three free parameters. However, as
pointed out in  \cite{Bzdak:2018uhv}, this distribution predicts a clear pattern for its factorial
cumulants\footnote{We note that STAR uses $C_n$ to denote cumulants, which we denote them by $\kappa_n$.}
\begin{eqnarray}
  \frac{C_{n+1}}{C_{n}} \simeq -17,
  \label{eq:fac_cum}
\end{eqnarray}
or, in other words, from order to order the factorial cumulants alternate in sign and increase 
in magnitude by a factor of $\sim 17$. This prediction of the two-component
model can and should be tested by extracting higher order factorial cumulants from the data. And, as
we shall elaborate below,
if the factorial cumulants up to order 7 or 8 also agree with the model prediction it is very likely
that we deal with a two-component distribution. If so, one may indeed have a system with two
``phases'' which can be further explored by measuring flow and other
observables, as discussed in more detail in  \cite{Bzdak:2018uhv}.

\begin{figure}[t]
\begin{center}
  \includegraphics[width=0.45\textwidth]{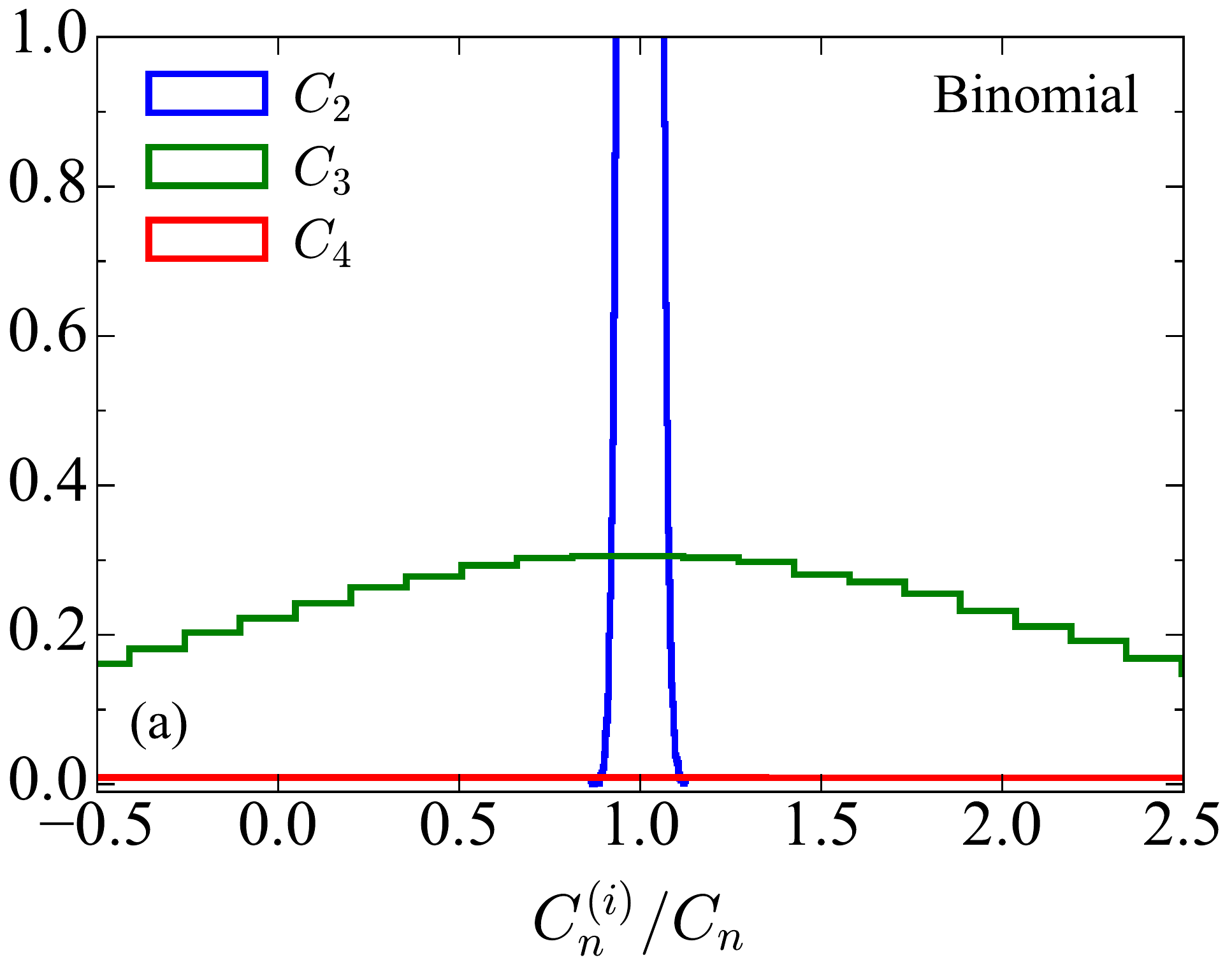}
  \hspace{0.05\textwidth}
  \includegraphics[width=0.435\textwidth]{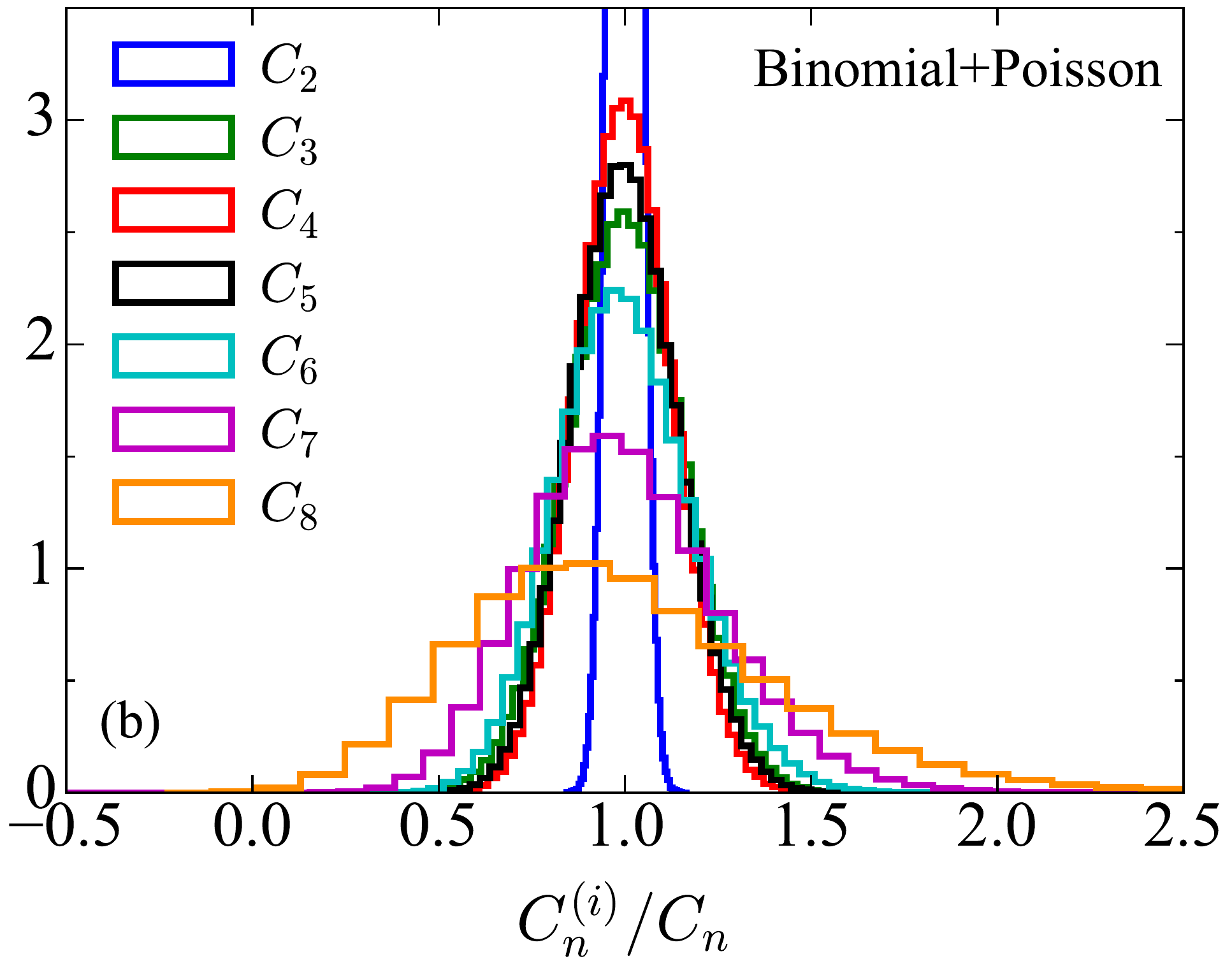}
\end{center}
\par
\vspace{-5mm}
\caption{Histogram (normalized to unity) of factorial cumulant, $C_{n}^{(i)}$, fluctuating from ``experiment'' 
  to ``experiment'', divided by a known (evaluated analytically) value, $C_{n}$, based on 150k events sampled from (a) a
  binomial distribution ($B=350$, $p=0.114$, $\langle N \rangle = pB \approx 40$) and (b) the distribution given 
  by Eq. (\ref{eq:two_component}).}
\label{fig:width_comparison}
\end{figure}

The  task of extracting higher order factorial cumulants is helped considerably by the fact that the above distribution,
Eq.~(\ref{eq:two_component}), is statistics friendly \cite{Bzdak:2018axe}. By this we mean that a 
surprisingly small number of events is needed to extract factorial cumulants of high order. This property is
due to the presence of the second component, even though it is rather small, $\alpha \simeq 0.3\%$, as
we demonstrate in Fig.~\ref{fig:width_comparison}. There we show the distribution of extracted values
for the factorial cumulants for ``experiments'' with $150k$ events,
approximately the number of events STAR has recorded for central collision at $\sqrt{s}=7.7\gev$. The width of
these distributions then give us the
expected statistical error for a measurement with $150k$ events. The left panel shows the
distribution if we just sample a binomial distribution, i.e. the distribution $P(N)$,
Eq.~(\ref{eq:two_component}), {\em without} the small component,  $\alpha =0$ (this distribution is plotted as
black dashed line in the right panel of Fig~\ref{fig:relative_error}). The right panel of Fig.~\ref{fig:width_comparison} shows the
distribution with the small component included. Clearly, the expected statistical error is much
smaller in the second case, even though we only have a very small admixture of the second component. This
is more clearly seen in the left panel of Fig.~\ref{fig:relative_error}, where we show the expected relative error for
factorial cumulants up to ninth order for a binomial, negative binomial distribution as well as for
the two-component distribution, Eq.~(\ref{eq:two_component}). Also shown, in magenta (open circles), are the
relative errors if we assume that the data are not efficiency corrected but still follow a two-component distribution (for details see~\cite{Bzdak:2018axe}). We note that the numerically
extracted statistical errors agree perfectly with those obtained from the so called delta method \cite{Luo:2014rea}.
More precisely, the predicted values of the factorial cumulants up to eighth order are (the first
four agree with the preliminary STAR data by construction):
\begin{eqnarray}
C_{5} &=& -307 \, (1\pm 0.31), \quad    C_{6} = 3085 \, (1\pm 0.41), \non
          C_{7} &=& -30155 \, (1\pm 0.61), \quad  C_{8} = 271492 \, (1\pm 1.06),
\label{eq:cn_predict_no_eff} 
\end{eqnarray}%
for efficiency uncorrected data and 
\begin{eqnarray}
C_{5} &=& -2645 \, (1\pm 0.14), \quad    C_{6} = 40900 \, (1\pm 0.18), \non
          C_{7} &=& -615135 \, (1\pm 0.26), \quad  C_{8} =8520220 \, (1\pm 0.42),
\label{eq:cn_predict_eff}                    
\end{eqnarray}%
for efficiency corrected data. The errors quoted here are only due to the sample size and do not
account for additional systematic  
uncertainties for example due to the efficiency corrections \cite{Luo:2014rea}.
Also, the central values of our
  prediction are based on the central values of the preliminary STAR data for the proton
  cumulants. In other words we did not do any error propagation here.

Given the small expected statistical errors, a measurement of factorial cumulants  up to the
eighth order should be possible if the data are efficiency corrected. If the factorial
cumulants extracted from the STAR measurement agree with the above predictions, it will provide
strong evidence that we are indeed dealing with a two-component multiplicity distribution.\footnote{We checked that the predicted values, Eq.~(\ref{eq:cn_predict_eff}), are practically
  impossible to result from a statistical fluctuation of a single binomial (or Poisson)
  distribution. For example, for a single binomial distribution with 150k events, the probability to
  obtain $C_{3}<0$, $C_{4}>0$, $C_{5}<0$ and $C_{6}>0$ is already about 0.04, and the probability to
  obtain numbers in the order of magnitude of Eq.~(\ref{eq:cn_predict_eff}) is practically
  zero. This can be easily understood since the absolute error of $C_{n}$ from a binomial
  distribution is close to $\sqrt{n!} \langle N \rangle^{n/2} / \sqrt{n_{\rm events}}$
  \cite{Bzdak:2018axe} which is significantly smaller than the numbers quoted in
  Eqs.~(\ref{eq:cn_predict_no_eff}) and (\ref{eq:cn_predict_eff}).}
This is demonstrated in the right panel of
Fig.~\ref{fig:relative_error}, where we show results of the Poisson-Charlier expansion for a  
probability distribution \cite{Charlier_0809.4155} at various orders. As discussed in more detail in
\cite{Bzdak:2018uhv}, the Poisson-Charlier expansion of order $n$ generates a probability
distribution based on the first $n$ factorial cumulants in such a way that it reproduces these first
$n$ factorial cumulants. If we use only the first four factorial cumulants generated from the two-component
distribution (which by construction agree with the STAR data),  the resulting Poisson Charlier
distribution (dot-dashed magenta line) does not agree with that of the two-component distribution (red
points). This demonstrates that, as already pointed out, four cumulants hardly constrain a
probability distribution. However, using the first six factorial cumulants (dashed green line), the
Poisson-Charlier distribution is already very close. And with the first eight factorial cumulants
the resulting Poisson-Charlier distribution (full blue line) is almost identical with that of the
two-component model (red points).
And, with the expected improved statistics from the second phase of the beam energy
scan, the distribution could be constrained even further.

Of course, even if a two-component distribution is experimentally confirmed, one still needs to rule out other
sources for such a distribution, such as a possible contamination of the data by events from a
different centrality class etc.

\section{Conclusions}
\label{sec:conclude}
In conclusion, we have shown that a two-component multiplicity distribution is statistics friendly
in the sense that factorial cumulants of rather high order may be extracted even with limited
statistics. This allows to test and confirm {\em right now} the hypothesis of a two-component model of
Ref.~\cite{Bzdak:2018uhv} with the presently available statistics of the STAR measurement from BES-I.
If this hypothesis is confirmed, and, if 
any possible experimental effects and backgrounds have been ruled out, we may actually have a first
glimpse at the QCD phase transition. This can then be further tested by measuring other observables such as flow for
events with small and large number of protons within the same centrality class.

\section{Acknowledgements}
A.B. was partially supported by the Ministry of Science and Higher Education, and by the National Science Centre, Grant No. 2018/30/Q/ST2/00101.
V.K. and D.O. were supported by the U.S. Department of Energy, Office
of Science, Office of Nuclear Physics, under contract number DE-AC02-05CH11231.  D.O. also
received support within the framework of the Beam Energy Scan Theory (BEST) Topical Collaboration. JS thanks the Samson AG and the BMBF through the ErUM-Data project for funding.
The computational resources where in part provided by the LOEWE Frankfurt Center for Scientific
Computing (LOEWE-CSC).











\end{document}